\documentclass[pdflatex,sn-mathphys-num]{sn-jnl}
\usepackage{graphicx}%
\usepackage{multirow}%
\usepackage{amsmath,amssymb,amsfonts}%
\usepackage{amsthm}%
\usepackage{mathrsfs}%
\usepackage[title]{appendix}%
\usepackage{xcolor}%
\usepackage{textcomp}%
\usepackage{manyfoot}%
\usepackage{booktabs}%
\usepackage{algorithm}%
\usepackage{algorithmicx}%
\usepackage{algpseudocode}%
\usepackage{listings}%

\begin{document}




\title[Charged Rotating Black Hole and the First Law]{Charged Rotating Black Hole and the First Law}

\author*{{\fnm S. D.} \sur{Campos}}\email{sergiodc@ufscar.br}

\affil*{\orgdiv{Department of Physics, Chemistry, and Mathematics}, \orgname{Federal University of S\~ao Carlos}, \orgaddress{{Rodovia Jo\~ao Leme dos Santos, km 110}, \city{Sorocaba}, \postcode{CEP 18052780}, \state{S\~ao Paulo}, \country{Brazil}}}

\abstract{
The thermodynamic properties of black holes have been extensively studied through analogies with classical systems, revealing fundamental connections between gravitation, entropy, and quantum mechanics. In this work, we extend the thermodynamic framework of black holes by incorporating charge and analyzing its role in entropy production. Using an analogy with charged rotating soap bubbles, we demonstrate that charge contributes to the total angular momentum and affects the entropy-event horizon relationship. By applying the Gouy-Stodola theorem, we establish a consistent thermodynamic formulation for charged black holes, showing that the first law of thermodynamics remains valid in this context. Furthermore, we explore the behavior of the partition function from the perspective of a distant observer, revealing that charge effects diminish with increasing distance. These findings reinforce the thermodynamic interpretation of black holes and provide insights into the interplay between charge, rotation, and entropy in gravitational systems.}






\maketitle
\section{Introduction}

As well established, the Schwarzschild solution describes static black holes \cite{K.Schwarzschild.Sitzungsber.Preuss.Akad.Wiss.Berlin.1916.189.1916}, while the Reissner-Nordström solution accounts for non-rotating charged black holes, both established by 1918 \cite{H.Reissner.Ann.Phys.59.106.1916,G.Nordstrom.Proc.Kon.Ned.Akad.Wet.20.1238.1918}.  Nonetheless, the comprehension of the black hole (BH) event horizon was only achieved in 1958 when Finkelstein \cite{D.Finkelstein.Phys.Rev.110.965.1958} clarified the nature of the BH event horizon, defining it as a gravitational membrane that separates the interior from the exterior. Subsequently, the theoretically significant instances of non-charged rotating black holes, known as the Kerr solution, emerged in 1963 \cite{R.P.Kerr.Phys.Rev.Lett.11.237.1963}, followed by the charged rotating black holes, referred to as the Kerr-Newman solution, in 1965 \cite{E.T.Newman.J.Math.Phys.6.918.1965}.

More than 50 years ago, Bekenstein (see, for example, Refs. \cite{j.d.bekenstein.lett.nuovo.cimento.4.15.737.1972,J.D.Bekenstein.Phys.Rev.D7.2333.1973,j.d.bekenstein.lett.nuovo.cimento.9.11.467.1974}) utilized a straightforward yet effective analogy to demonstrate a potential relationship between entropy and the event horizon area in black holes. Thereafter, Hawking (for example, Refs. \cite{J.M.Bardeen.B.Carter.S.W.Hawking.Commun.Math.Phys.31.161.1973,s.w.hawking.1974,S.W.Hawking.Commun.Math.Phys.43.199.1975,s.w.hawking.1976}) incorporated the formalism of quantum mechanics into Bekenstein's proposal and, remarkably, arrived at identical conclusions. This suggests a resemblance between the laws of thermodynamics and the principles governing BH behavior.

During the golden era of BH research (1960–1980), one key discovery was the entropy-event horizon area law. This law extends thermodynamics beyond conventional matter, as BHs represent an extreme state that challenges fundamental physics. The analogies drawn between the principles of thermodynamics and the physics governing black holes can be regarded as an initial step toward formulating comprehensive general laws of physics.

Hawking, Page, and Pope \cite{S.W.Hawking.D.N.Page.C.N.Pope.Phys.Lett.86B.175.1979,S.W.Hawking.D.N.Page.C.N.Pope.Nucl.Phys.B.170.2.283.1980} proposed the idea of a gravitational bubble to account for what they term as virtual black holes, which are entities capable of emerging from the vacuum and subsequently vanishing \cite{S.W.Hawking.D.N.Page.C.N.Pope.Nucl.Phys.B.170.2.283.1980}. Since then, analogies between soap bubbles and BH were explored over the years (see, for example, Refs. \cite{d.m.eardley.phys.rev.d.57.1998,d.m.eardley.phys.rev.d.66.2002,j.l.jaramillo.phys.rev.d.89.2014}). Recently, a non-charged rotating black hole was treated as a gravitational bubble \cite{campos.longaresi.ijmpd.2025}, allowing the obtaining of a thermodynamic framework for the BH. Through implementing the Gouy-Stodola theorem, a gravitational tension enabling a coherent thermodynamic treatment for the entropy-event horizon law was established. 

In the present work, the charge of the BH, as well as mass and angular momentum, is also considered to explain the entropy-event horizon law for the BH. The objective of the present analysis is to demonstrate, employing the analogy between the soap bubble and the gravitational bubble generated by the BH, and utilizing the Gouy-Stodola theorem, that the first law of thermodynamics is compatible with the thermodynamic laws for black holes \cite{J.M.Bardeen.B.Carter.S.W.Hawking.Commun.Math.Phys.31.161.1973}.

This paper is organized as follows. In Section \ref{sec:charge}, one compares the behavior of a charged rotating soap bubble and a charged rotating BH. Both objects present the same behavior for the charge stored by the electromagnetic angular momentum. Section \ref{sec:chargerot} presents the results for the first law considering the charge. Section \ref{sec:partfunc} discusses the behavior of the partition function for a distant observer, while Section \ref{sec:disc} provides some general remarks. 

\section{Charged Rotating}\label{sec:charge}

This section examines the similar behavior between the charge associated with the electromagnetic angular momentum of a spherical soap bubble and that of a charged rotating BH. In both scenarios, the magnitude of the stored charge diminishes with increasing distance from the physical entity.

\subsection{Soap Bubbles}

The theoretical framework regarding the equilibrium of charged water droplets has been established since 1882, following Lord Rayleigh \cite{L.Rayleigh.Philos.Mag.14.184.1882}. The introduction of an electric charge, irrespective of its sign, increases the droplet radius due to the effect of electrical repulsion (its pressure drops due to outward electric pressure). Naturally, the upper limit of the droplet radius depends on the chemical composition of the substances used to generate it. 

Conventionally, it is assumed that a spherical soap bubble exhibits a uniform charge distribution; however, this assumption may not hold when the gravitational potential influence is considered. The presence of gravitational forces modifies the thickness of the membrane towards the center of attraction, resulting in the membrane being thinner on the top than the bottom \cite{A.Pelesz.P.Zylka.Exp.in.Fluids.61.241.2020,Y.D.Afanasyev.G.T.Andrews.C.G.Deacon.Am.J.Phys.79.1079.2011}. Nevertheless, for approximation, I shall assume a uniform charge distribution. This assumption allows the consideration of the charged spherical soap bubble as a spherically charged shell possessing a surface tension. 

The balance between electrostatic and excess pressure in an incompressible spherical soap bubble, in thermodynamic equilibrium, characterized by a diameter $r$ and surface tension $\sigma$ is given by \cite{isenberg_book}
\begin{eqnarray}
    \frac{\epsilon_0 E^2}{2}=\frac{4\sigma}{r},
\end{eqnarray} 
with $\epsilon_0$ denoting the vacuum permittivity. It is interesting that, in the context of a BH, the surface tension $\sigma$ can be associated with the surface gravity $\kappa$ within the framework of a Young-Laplace law for BH horizons \cite{j.l.jaramillo.phys.rev.d.89.2014}, reinforcing the analogy between BH and soap bubbles used here and elsewhere.

As well-established, the total charge on a spherical shell is written as $q=4\pi r^2\epsilon_0 E$, resulting in the expression for the maximum charge sustainable by the bubble given as \cite{L.Rayleigh.Philos.Mag.14.184.1882,A.D.Bonhommeau.J.ofChem.Phys.146.12.124314.2017} 
\begin{eqnarray}\label{eq:chargedbubble}
    q_r=8\pi\sqrt{2\epsilon_0\sigma r^3},
\end{eqnarray}
indicating the bubble becomes unstable under the condition of $q>q_r$. When the charged spherical soap bubble of mass $m$ rotates with non-constant angular velocity $\omega=\omega(t)$ ($\hat{r}$ direction) and moment of inertia $I$, one should add to the usual mechanical angular momentum 
\begin{eqnarray}
    L_{m}=I\omega = \frac{2mr^2\omega}{3},
\end{eqnarray}
the electromagnetic angular momentum of the spherical shell given by \cite{griffths_book}
\begin{eqnarray}
    L_{e}=\epsilon_0\int d^3r[\vec{r}\times(\vec{E}\times\vec{B})]=\frac{q^2r\omega}{18\pi\epsilon_0},
\end{eqnarray}
considering the electric and magnetic fields \cite{A.S.Castro.Am.J.Phys.59.2.180.1991}
\begin{eqnarray}
\nonumber   \vec{E}=\frac{q}{4\pi\epsilon_0 r^2}\hat{r}~~ \mathrm{and}~~\vec{B}=\frac{q r^2}{12\pi\epsilon_0 r'^3}\bigl[3\vec{r}(\vec{r}\omega)-\omega \bigr],
\end{eqnarray}
where $r'>r$.  The spherical shell radiates electromagnetic radiation due to the non-constant angular velocity, which takes angular momentum away from the system. Subsequently, the interaction between the electric and magnetic fields gives rise to an intrinsic torque, resulting in an electromagnetic angular momentum \cite{griffths_book}. The total angular momentum of this rotating charged spherical soap bubble is
\begin{eqnarray}\label{eq:momang_1}
    L_T=L_m+L_e=\bigl(I_m+I_e\bigr)\omega=\left(\frac{2mr^2}{3}+\frac{q^2r}{18\pi\epsilon_0}\right)\omega.
\end{eqnarray}

Reorganizing equation \eqref{eq:momang_1} in terms of charge $q$, one can write this result depending on $q_r$ as
\begin{eqnarray}\label{eq:df1}
    q=\frac{3q_r}{8r}\sqrt{\frac{m}{\pi\sigma}\left(\frac{I_T}{m}-\frac{2r^2}{3}\right)\omega}=\frac{3q_r}{8r}\sqrt{\frac{L_T - L_m}{\pi\sigma}},
\end{eqnarray}
where $I_T$ is the total moment of inertia with the constraint $I_T\geq I_m=(2/3)mr^2$. It should be noted that $L_T \rightarrow L_m$ implies $q\rightarrow 0$, indicating the absence of any charge stored by electromagnetic angular momentum. One can write the $L_T-Lm$ as 
\begin{eqnarray}
\left(1-\frac{L_m}{L_T}\right)L_T=\delta L_T<L_T,
\end{eqnarray}
implying $\delta<1$. Observe that the very slow-rotating soap bubble case implies the extreme condition $L_m<\!\!<L_T$: The total angular momentum is essentially due to the charge stored by electromagnetic angular momentum. On the other hand, the fast-rotating case means $L_e<\!\!<L_m\approx L_T$. Using the above result, one can rewrite equation \eqref{eq:df1} as
\begin{eqnarray}\label{eq:chargee}
    q=\frac{3q_r}{8r}\sqrt{\frac{\delta L_T}{\pi\sigma}},
\end{eqnarray} 
corresponding to the total charge stored by the electromagnetic angular momentum outside the soap bubble. This value is constrained by result \eqref{eq:chargedbubble}, implying the parameter $\delta L_T$ possesses a constraint given by $q\lesssim q_r$. Then, using results \eqref{eq:chargedbubble} and \eqref{eq:chargee}, one writes
\begin{eqnarray}\label{eq:df2}
\delta L_T\lesssim \frac{64\pi\sigma r^2}{9}.
\end{eqnarray}

The above constraint allows us to assume that there exists $r<d(\delta)=d\in \mathbb{R}$ such that
\begin{eqnarray}\label{eq:props}
    \delta L_T=\frac{64\pi\sigma}{9} \left(\frac{r}{d}\right)^2.
\end{eqnarray}

A possible interpretation for this parameter is related to the distance we are from the spherical soap bubble: For each spherical soap bubble with radius $r$, there exists a spherical shell characterized by a radius $r<d$, from which one measures the charge stored by the electromagnetic angular momentum. For a distant observer (in a convenient reference frame), the electromagnetic angular momentum vanishes as $d\rightarrow\infty$. Then, using the ansatz \eqref{eq:props}, one rewrites equation \eqref{eq:chargee} as
\begin{eqnarray}\label{eq:chargee2}
    q=\frac{3q_r}{8r}\sqrt{\frac{\delta L_T}{\pi\sigma}}=\frac{q_r}{d}.
\end{eqnarray} 

\begin{figure}[h!]
    \centering
    \includegraphics[width=0.6\linewidth]{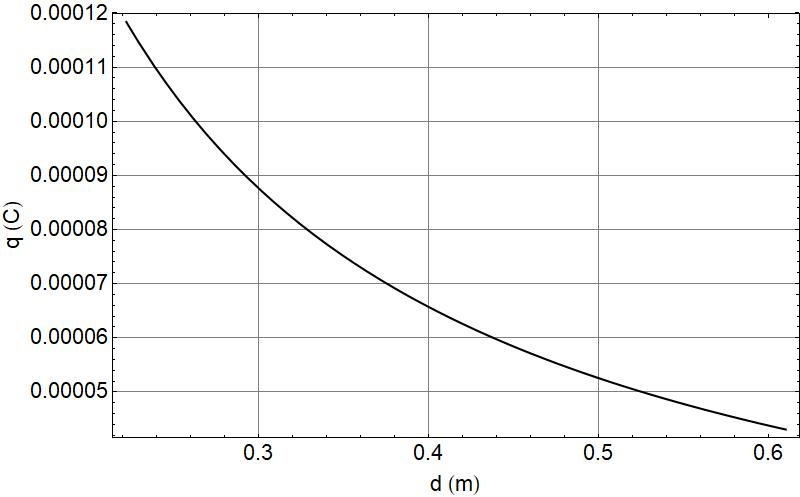}
    \caption{\label{fig:bubble_2}Charge stored by the electromagnetic angular momentum for a spherical soap bubble, where one uses $r=0.112$ m, and $\sigma=34$ mF/m from Ref. \cite{A.Pelesz.P.Zylka.Exp.in.Fluids.61.241.2020}. Observe that the radius $r=0.6$ m corresponds to a huge soap bubble.}
\end{figure}

Then, the stored charge corresponds to a small fraction of the limiting charge $q_r$. Therefore, when $d$ grows, the charge stored in the electromagnetic angular momentum diminishes, as shown in Figure \ref{fig:bubble_2}.

\subsection{Black Holes}

Within the framework of a BH, the extremal condition restricts the indefinite accumulation of charge by the object. Upon attaining its extremality limit $Q_{max}$, any further attempt to increase the charge results in either: 1) the repulsion of the incoming charged particle by the black hole's electrostatic field, or 2) if is possible to have $Q>Q_{max}$, then the event horizon would disappear, leading to the formation of a naked singularity, which is forbidden by the cosmic censorship conjecture.  

A BH with angular velocity also presents electromagnetic angular momentum as well as intrinsic angular momentum (the so-called intrinsic spin). This subject is not new and can be viewed in detail in Refs.  \cite{D.Garfinkle.S.J.Rey.Phys.Lett.B.257.158.1991,J.H.Kim.S.H.Moon.JHEP.09.088.2007,C.Bunster.M.Henneaux.cecs.phy}. The solution presented in Ref. \cite{D.Garfinkle.S.J.Rey.Phys.Lett.B.257.158.1991}, based on very general conditions, assumes the BH and a test particle with charge $e$ form an axisymmetric stationary system with Killing fields $(\partial/\partial t)^a$ and $(\partial/\partial\varphi)^a)$, where $(-t,r,\theta,\varphi)$. Using the electromagnetic field tensor,  the Einstein-Maxwell, and the Einstein equations, they found that the electromagnetic angular momentum depends on the distance $b$ between the particle with charge $e$ and the BH with charge $Q$ as (for a sufficiently large $b$ and adopting $c=\hbar=G=k_B=1$) \cite{D.Garfinkle.S.J.Rey.Phys.Lett.B.257.158.1991}
\begin{eqnarray}\label{eq:garfinkle}
    L_e=eQ\left[1- \frac{2r_H}{b^2}\left(\frac{M^2-Q^2}{2r_H-M} \right) \right],
\end{eqnarray}
where the event horizon is the one for the Reissner-Nordstr\"om BH
\begin{eqnarray}
    r_H=M+\sqrt{M^2-Q^2}= M\bigl(1+\sqrt{1-y^2} \bigr),
\end{eqnarray}
where $y=Q/M$. In the approach of Garfinkle and Rey \cite{D.Garfinkle.S.J.Rey.Phys.Lett.B.257.158.1991}, they consider the intrinsic spin $S$ of the BH manifests itself within the event horizon, resulting in the total angular momentum of the BH being given by \cite{D.Garfinkle.S.J.Rey.Phys.Lett.B.257.158.1991}
\begin{eqnarray}
    J=S+L_e.
\end{eqnarray}

To estimate the charge stored by the electromagnetic angular momentum of the BH, one restricts the problem to the following conditions: (1) $b>>r_H$ (a distant charge) and (2) $Q/M<1$ (a small charge compared to $M$). In this case, one can use the result \eqref{eq:garfinkle}, allowing us to consider the first-order series expansion for $y$
\begin{eqnarray}
    r_H\approx M\left(2-\frac{1}{2}y^2 \right).
\end{eqnarray}

As performed for the spherical soap bubble, the total angular momentum $J$ is the sum of the intrinsic spin $S$ of the BH with the electromagnetic angular momentum shown in result \eqref{eq:garfinkle}. Considering $(1-S/J)J=\Delta J$, one can write
\begin{eqnarray}\label{eq:df3}
   \Delta J=eQ\left[1- \frac{2r_H}{b^2}\left(\frac{M^2-Q^2}{2r_H-M} \right)\right].
\end{eqnarray}

\begin{figure}[h!]
    \centering
    \includegraphics[width=0.6\linewidth]{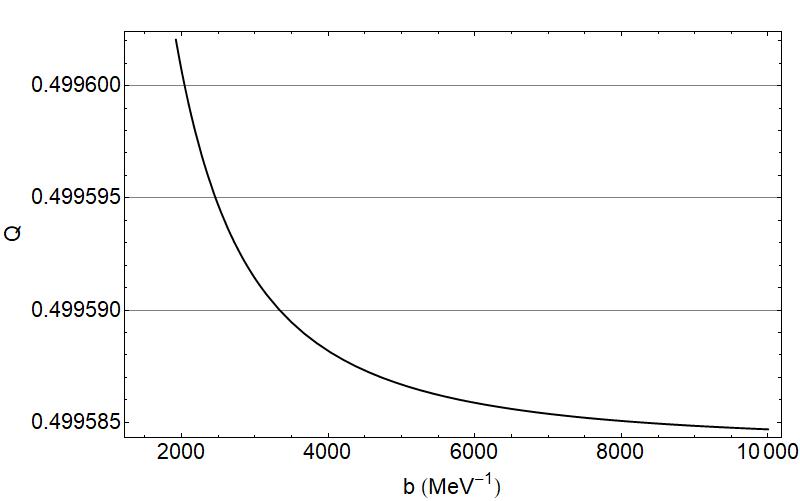}
\caption{\label{fig:charge_bh}Charge stored by the electromagnetic momentum angular for the BH case for $M=10$ MeV and $\delta J=0.5$. It should be noted that, analogous to the case of the spherical soap bubble, the charge diminishes with the increase of $b$.}
\end{figure}

Retaining only the quadratic terms of $y$, one has from equation \eqref{eq:df3} the following second-order equation for the charge-mass ratio 
\begin{eqnarray}
    y^2+\frac{Me}{\Delta J b^2}\bigl(3b^2-4M^2 \bigr)y-3=0,
\end{eqnarray}
whose only solution for $Q/M$, taking into account the conditions stated above, is written as
\begin{eqnarray}
    \frac{Q}{M}=\frac{-Me(3b^2-4M^2)+\sqrt{M^2e^2\bigl(3b^2- 4M^2\bigr)^2+12(\Delta J)^2b^2}}{2\Delta J b^2},
\end{eqnarray}
indicating the charge-mass relation decreases as $b$ grows. Observe that for $b\rightarrow \infty$, one has $Q/M\rightarrow 0$, as expected from the results above. However, considering $M<\!\!<b$ and $1<M/\Delta J$, and taking into account only the first-order term in the series expansion for $\sqrt{M^2e^2\bigl(3b^2- 4M^2\bigr)^2+12(\Delta J)^2b^2}$, one has
\begin{eqnarray}
    \frac{(\Delta J)^2}{2(Me)^2}<1,
\end{eqnarray}
implying $Q/M\rightarrow \Delta J/M e$ (or $Q\rightarrow \Delta J/ e$). Figure \ref{fig:charge_bh} shows the behavior of $Q$ according to $b$ for fixed $M$ and $\Delta J$. As expected, the charge stored by the electromagnetic angular momentum decreases as $b$ grows. It is interesting to note that the Hawking radiation, whatever it means, has its origin in quantum effects incorporated into the classical description of a BH \cite{s.w.hawking.1975}. For the charged radiation, it is reasonable to suppose that it can be (at least part of it) stored by the electromagnetic angular momentum, being negligible for an observer far from the BH (and in a far future), as shown here. However, part of these charged particles stored within the electromagnetic angular momentum may be captured by the gravitational field of the BH, eventually entering through the event horizon, thereby extending the lifetime of the BH.

In light of the findings from the preceding subsections, one notes that for both the spherical soap bubble and BH, the stored charge decreases as the electromagnetic angular momentum diminishes with distance. In the case of the soap bubble, this is postulated as an ad hoc assumption due to the inherent capacity to store a maximum charge. Conversely, in the BH scenario, this decrease is obtained as a theoretical outcome. This analogy, as discussed in Ref. \cite{campos.longaresi.ijmpd.2025}, highlights the conceptual similarity between the soap bubble and the BH.

\section{First Law for a Charged Rotating Black Hole}\label{sec:chargerot}

Considering a BH with fixed event horizon radius $R_f$, possessing total charge $Q$, mass $M$, and total angular momentum $J$ in a fixed background spacetime, the charge in the gravitational surface is given by event horizon radius in the Kerr-Newman metric written as
\begin{eqnarray}\label{eq:charge_1}
    Q=\sqrt{2MR_f-\bigl(R_f^2+a^2\cos^2\theta\bigr)},
\end{eqnarray}
where $a=J/M$ is the Kerr parameter, and the total charge $Q$ is real-valued, considering the constraint 
\begin{eqnarray}\label{eq:constraint}
    (J/M)^2\cos^2\theta+R_f^2\leq 2MR_f.
\end{eqnarray}

\begin{figure}[h!]
    \centering
    \includegraphics[width=0.6\linewidth]{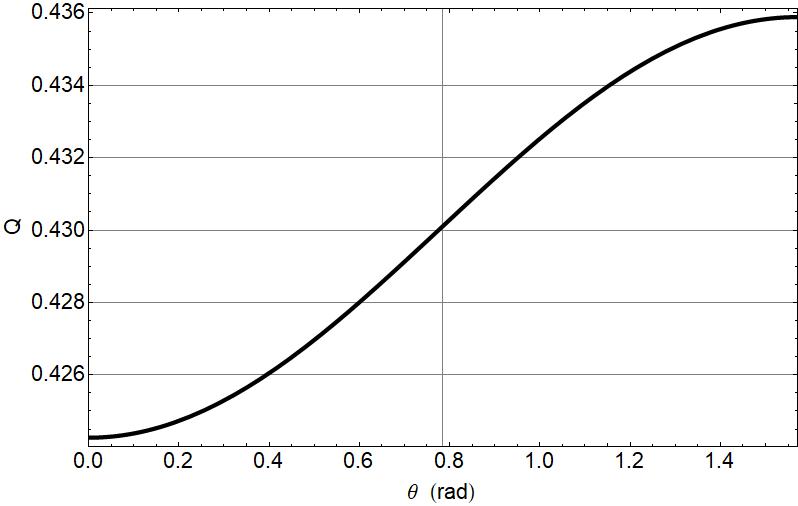}
    \caption{\label{fig:chargevariation}Charge variation according to $\theta$ for $J=0.1$, $R=1$ MeV$^{-1}$ and $M=10$ MeV, and respecting the constraint \eqref{eq:constraint}.}
\end{figure}

The charge variation according to $\theta$ and respecting the constraint \eqref{eq:constraint} is shown in Figure \ref{fig:chargevariation} for fixed $R_f$, and, as expected, $dQ/d\theta=0$ has its maximum values at $\theta=\pm\pi/2$. Then, the total charge $Q$ has a small non-homogeneous distribution along the event horizon surface. 

Consider now that $E$ corresponds to the total energy and $V$ (depending on the event horizon) is the physical volume enclosed by the BH. The term ``volume'' means here spatial volume, being not slicing invariant since it is spacetime dependent \cite{m.k.parikh.phys.rev.d.73.2006}. In some conditions, a vector volume can be defined in spacetime, allowing the definition of an invariant conserved volume \cite{w.ballik.klake.phys.rev.d.88.2013}. It is interesting to note that in the work of Parick \cite{m.k.parikh.phys.rev.d.73.2006}, under specific conditions, the volume for the BH can be constant in time, while in the work of Christodoulou and Rovelli \cite{christodoulou.rovelli.2015}, the volume of the BH is time-dependent. For a recent review on this subject, please see Ref. \cite{ali.t.liu.2024}, and for a pedagogical introduction on the subject, please see Ref. \cite{dinunno.matzner.2010}. Here, I assume the BH volume can be defined in a way that the thermodynamic term $PdV$ is non-zero, similarly to the thermodynamic BH volume proposed by Jaramillo \cite{j.l.jaramillo.phys.rev.d.89.2014}, where surface tension is used to define the intensive variable related to the BH volume (the extensive variable). Then, using the Gouy-Stodola theorem and the analogy between the soap bubble and BH behavior, it has been shown for the (non-charged) rotating BH that the first law of thermodynamics \cite{campos.longaresi.ijmpd.2025}
\begin{eqnarray}\label{eq:firstlaw1}
    dE= TdS - pdV,
\end{eqnarray}
is equivalent to the first law of thermodynamics written for a BH as \cite{J.M.Bardeen.B.Carter.S.W.Hawking.Commun.Math.Phys.31.161.1973}
\begin{eqnarray}\label{eq:bardeen1}
    dM=\frac{\kappa}{8\pi}dA + \Omega dJ,
\end{eqnarray}
where $\Omega=d\theta/dt$ is the BH angular velocity, and $dA$ is the event horizon area proportional to the BH entropy. 

Using the usual analogy of a charged rotating BH, the first law of thermodynamics is written as \cite{J.M.Bardeen.B.Carter.S.W.Hawking.Commun.Math.Phys.31.161.1973}
\begin{eqnarray}\label{eq:tds1}
    dM=\frac{\kappa}{8\pi}dA + \Omega dJ+\Phi dQ,
\end{eqnarray}
being $\Phi$ is the electric potential due to the charge $e$. As is well known, the first law of thermodynamics \eqref{eq:firstlaw1} can also be written taking into account the chemical potential $\mu$ of the particles (that interact with the BH) as 
\begin{eqnarray}\label{eq:firstlaw10}
    dE= TdS - pdV+ \mu dN,
\end{eqnarray}
where $dN$ represents the number of particles attracted towards the BH gravitational surface. It is important to note that neither $dN$ nor $dQ$ are genuine infinitesimal quantities; rather, they are discrete integer values, albeit small in magnitude. 

Considering a BH, one notices that the last term in the rhs of equation \eqref{eq:firstlaw10} represents the useful work done by the surroundings on the BH gravitational surface, written in terms of chemical potential as \cite{Ma_book} 
\begin{eqnarray}
    dW=\mu dN,
\end{eqnarray}
and for a system in thermodynamic equilibrium, the chemical potential is constant, independent of the number of phases, and the particles flow from high to low chemical potentials (from higher to lower concentration), regulating the particle transfer between two systems in contact \cite{kittel_book}. 

For a BH, whose gravitational attraction is the dominant physical effect, one expects the chemical potential to have a vanishing contribution to explain the transfer of particles from the surroundings to the BH interior. Therefore, the BH charge can be, in general, disregarded. Nevertheless, considering the gravitational surface as a physical entity that separates the exterior from the BH interior, the electrochemical potential $\mu_{tot}$ depends on external and internal contributions \cite{D.Walz.S.R.Kaplan.Bioelectrochemistry.Bioenergetics.28.5.1992}
\begin{eqnarray}
    \mu_{tot}=\mu_{int}+\mu_{ext},
\end{eqnarray}
where $\mu_{int}$, in the present case, depends on the (inaccessible) internal features of the BH, while the second term on the rhs represents the contribution coming from the electromagnetic field \cite{D.Walz.S.R.Kaplan.Bioelectrochemistry.Bioenergetics.28.5.1992}
\begin{eqnarray}
    \mu_{ext}=\lambda F\Phi'
\end{eqnarray}
where $F$ corresponds to the ratio of the charge $e$ to the amount of particles $dN$, $\lambda$ corresponds to the charge number of the ion, and $\Phi'$ is the electrostatic potential felt by $e$. 

Of course, one can generalize the above result considering different types of particles with total charge $q$. In this case, one has
\begin{eqnarray}\label{eq:sum1}
   \sum_{i}\mu_idN_i=\Phi' \sum_{i}\lambda_i q\frac{dN_i}{dN},
\end{eqnarray}
where $dN_i$ represents the total number of particles of type-$i$. Notice that each term of the sum \eqref{eq:sum1} represents a fraction of the total charge $q$ corresponding to the type-$i$
\begin{eqnarray}
    q\frac{dN_i}{dN}=dq_i,
\end{eqnarray}
implying the sum is equal to the total charge $q$
\begin{eqnarray}\label{eq:sum2}
   \sum_{i}\lambda_i dq_i=q.
\end{eqnarray}

On the other hand, one can reorganize \eqref{eq:sum1} in a more convenient way for our purposes as
\begin{eqnarray}\label{eq:sum3}
   \sum_{i}\mu_idN_i=\Phi \sum_{i}\lambda_i Q\frac{dN_i}{dN},
\end{eqnarray}
where $\Phi=(k/r)q$ is due to the total charge $q$. In this case, each term of the sum represents an integer part of the total charge $Q$ of the BH
\begin{eqnarray}
    Q\frac{dN_i}{dN}=dQ_i.
\end{eqnarray}

Notice, now, that the sum on the rhs of the equation \eqref{eq:sum3} is not equal to the total charge $Q$, since each $\lambda_i$ is related to the type $i$ whose total charge is $q$ (not $Q$). Then, this sum represents a small integer part of $dQ$
\begin{eqnarray}
    \sum_{i} \lambda_i dQ_i=dQ.
\end{eqnarray}

Therefore, one has
\begin{eqnarray}\label{eq:finalddd}
    dW=\sum_{i}\mu_idN_i=\Phi' q=\Phi dQ.
\end{eqnarray}

The Gouy-Stodola theorem states that the entropy yielded, for $\dot{T}=0$, is given by \cite{G.Gouy.J.dePhys.8.35.1899,A.Stodola.Zeitschr.d.Verein.Deutscher.Ingenieure.32.1086.1898}
\begin{eqnarray}
    Td\dot{S}=d\dot{W}_r-d\dot{W}_i,
\end{eqnarray}
where $d\dot{W}_r$ corresponds to the time variation of the work due to conservative (reversible) processes, and $d\dot{W}_i$ is due to non-conservative (irreversible) ones. Of course, one can write   
\begin{eqnarray}
    TdS=dW_r-dW_i,
\end{eqnarray}
implying the difference above is the sum of the useful work $dW$ given by \eqref{eq:finalddd} and the other contributions from the $dE$ and $pdV$ \cite{campos.longaresi.ijmpd.2025}. Then, it allows us to write the sum of the thermodynamic quantities $dE$, $pdV$, and $\sum_{i}\mu_i dN_i$ as equal to the difference between the reversible and irreversible infinitesimal work  
\begin{eqnarray}\label{eq:finalfff}
    TdS=dE+pdV-\sum_{i}\mu_i dN_i=dW_r-dW_i.
\end{eqnarray}

Therefore, we need to add the result \eqref{eq:finalddd} to the (partial) first law given by equation \eqref{eq:bardeen1} to write
\begin{eqnarray}\label{eq:finalcvc}
    TdS=dE+pdV-\sum_{i}\mu_i dN_i\approx dM-\Omega dJ-\Phi dQ,
\end{eqnarray}
corresponding to the well-known result for charged rotating black holes \cite{J.M.Bardeen.B.Carter.S.W.Hawking.Commun.Math.Phys.31.161.1973}. The heuristic framework proposed in this study, drawing on the analogy between the charged rotating soap bubble and the charged rotating BH, alongside the Gouy-Stodola theorem, provides an interesting basis for deriving the first law of thermodynamics.


\section{Partition Function}\label{sec:partfunc}

As well known, the partition function is the tool that allows us to jump from a statistical to a thermodynamical description of a given system. Since the approach performed here has a thermodynamical origin, it is necessary, then, to ask for the statistical description derived from it since from the partition function one can obtain, for instance, the free energy of a system.

It is important to point out that, from a path integral formalism for the partition function, Gibbons and Hawking were able to obtain insights about the thermodynamic and statistical behavior of a BH \cite{gibbons.hawking}. In the present case, however, we are able to obtain the grand canonical partition function $Z$ from the entropy \eqref{eq:finalcvc}. Actually, the entropy of the BH can be written from the usual thermodynamic approach as 
\begin{eqnarray}
   S=\ln Z + T \left(\frac{\partial \ln Z}{\partial T}\right)_V.
\end{eqnarray}

Considering $\dot{T}= 0$ and neglecting high-order terms (taking $\partial \ln Z/\partial T=0$, for constant $V$), one has from result \eqref{eq:finalcvc} the non-normalized partition function
\begin{eqnarray}\label{eq:sdc}
    Z\approx e^{\Sigma T^{-1}},
\end{eqnarray}
where $\Sigma=(M-\Omega J+\Phi Q)$ can be viewed as the thermodynamic potential representing the energy available in the system. 

Assuming $r=b$ in the electrostatic potential, then the only variable of interest is the distance $b$ between the observer and the BH. It is not difficult to see that the partition function $Z$, for a fixed $T$, tends to a constant value as $b$ increases above some critical value $b_c$ since the charge stored by the electromagnetic angular momentum tends to a (small) limiting value. Hence, an observer far from the BH cannot distinguish between the energy states of the system. For $b_c<b$, the charge seems not to be a relevant feature to describe a BH. Certainly, this is a foreseeable scenario because the key attribute of a BH regards the gravitational effects that arise from its presence in a particular region of space.

To corroborate this, observe that the partition function obtained by Gibbons and Hawking \cite{gibbons.hawking} for the quantization of a field $\phi$ depends on the action $I(g_0,\phi_0)$ 
\begin{eqnarray}\label{eq:gibbons}
    \ln Z=iI(g_0,\phi_0)+(\mathrm{gravitational}+\mathrm{matter})~\mathrm{terms},
\end{eqnarray}
where $g_0$ and $\phi_0$ are background fields. Comparing the results \eqref{eq:gibbons} and \eqref{eq:sdc}, one observes that \eqref{eq:gibbons} tends to a constant value for $b_c<b$. This implies that the distant observer cannot differentiate the quantized field from the BH surrounding background. As the observer approaches the BH, it becomes reasonable to assume that the quantized field and the background can be differentiated. This issue is analogous to a scaling problem where a granular surface appears smooth to an observer far away from the surface.

Furthermore, for $b_c<b$ , one has $\dot{Z}\approx 0$, implying $\dot{S}\approx 0$ for a distant observer. Actually, the time derivative of equation \eqref{eq:sdc} can be written as 
\begin{eqnarray}
    \dot{S}e^{\Sigma T^{-1}}\approx 0,
\end{eqnarray}
implying $\dot{S}\approx 0$. In the present approach, this result indicates that the entropy production has only local relevance. Then, for a distant observer ($b_c<b$), the BH entropy seems to be a constant of motion. From the quantum mechanics viewpoint, this can be regarded as a consequence of the purity of the states describing the system. Then, for a distant observer, this result represents the absence of mixed states on the surface of a Bloch sphere of radius $b$. As $b$ decreases below $b_c$, mixed states inside the Bloch sphere of radius $b$ may become relevant, turning the entropy production into a significant quantity. 

Observe that the physical effects of the Hawking radiation are restricted to the inside of the Bloch sphere, turning its detection a real challenge since the observer far from the BH, according to the results presented here, only measures pure observable states of a quantum field (radiation). Moreover, in the first moment, the entanglement between the Hawking radiation and the radiation surrounding the BH also occurs inside the Bloch sphere.


\section{Discussion}\label{sec:disc}

The results presented in this work strengthen the connection between BH thermodynamics and classical thermodynamic principles, particularly through the analogy with charged rotating soap bubbles. By extending previous analyses \cite{campos.longaresi.ijmpd.2025} to incorporate charge, we demonstrate that the entropy-event horizon relationship remains consistent with the first law of thermodynamics, even in the presence of electric charge. 

The thermodynamic treatment of charge in rotating black holes is crucial for understanding Hawking radiation and BH evaporation. In particular, the role of electromagnetic angular momentum in storing charge suggests that some charged particles may be temporarily retained by the gravitational surface of the BH. This effect could influence the charge-dependent emission spectra of BH, affecting models of primordial BH evaporation and their potential contribution to cosmic-ray production. Furthermore, since charge can modify the BH specific heat and stability conditions, the results presented here may be relevant for studying BH phase transitions, particularly in the context of AdS space and extended thermodynamics.

A charged rotating BH could exist in environments where plasma interactions or magnetic fields lead to charge accumulation. The results presented here suggest that at large distances, the charge effects become negligible, meaning that observers detecting BH emissions (e.g., radio or X-ray signals from quasars) may only measure gravitational interactions, while local plasma dynamics near the event horizon could still be influenced by charge. This is particularly relevant for high-energy emissions from active galactic nuclei and the study of relativistic jets, where BH charge may play an indirect but measurable role. Additionally, as gravitational wave detectors become more sensitive, the impact of BH charge on gravitational wave signatures could be an avenue for observational constraints on BH charge.

In conclusion, the present work shows that charge does not violate BH thermodynamics but rather adds complexity to entropy dynamics. The results offer theoretical insights that could be tested in astrophysical observations and gravitational wave data, for example, strengthening the role of black holes as laboratories for studying the intersection of gravity, thermodynamics, and quantum theory, as recently done \cite{G.Bianconi.Phys.Rev.D.2025}.

\bmhead{Acknowledgements}

SDC thanks UFSCar for the financial support.



\end{document}